\documentclass[article,amsmath,amssymb,superscriptaddress,notitlepage]{revtex4-1}
\usepackage[utf8]{inputenc}
\usepackage[english]{babel}
\usepackage[titletoc,toc,title]{appendix}
\usepackage{amsmath}
\usepackage{amsfonts}
\usepackage{amssymb}
\usepackage[colorlinks=true,linkcolor=blue,citecolor=blue,urlcolor=blue]{hyperref}
\usepackage{graphicx}
\usepackage{enumerate}
\usepackage{csquotes}
\usepackage[caption=false]{subfig}
\usepackage{dsfont}
\usepackage{color}
\usepackage{bbold}
\usepackage{mathtools}
\usepackage{tensor}
\usepackage[export]{adjustbox}
\usepackage{hyperref}

\usepackage{epigraph}
\begin{document}

	\title{Commutative Factorization of Nonlinear Second-Order Differential Equations: Theory and Applications}
	
	\author{G. Gonz\'alez$^\dagger$}

\email{gabriel.gonzalez@academicos.udg.mx}
	
		%\homepage{http://xxx.yyy.zz/name}

	\affiliation{
$^\star$~Departamento de Ciencias Básicas y Aplicadas, Universidad de Guadalajara, CUTonalá Avenida Nuevo Periférico 555, Ejido San José Tateposco, Tonalá, 45425, Mexico\\}
%$^\diamond$~Department of Mathematics, Embry-Riddle Aeronautical University, Daytona Beach, FL 32114-3900, USA}

\begin{abstract}
 This work presents a generalization of the commutative factorization framework for a broad class of second-order nonlinear ordinary differential equations. The main objective was to extend the commutative factorization procedure in order to use the Riccati-Bernoulli equation and the Bäcklund transformation. This provides a systematic route for constructing both particular and general solutions of the original second-order nonlinear equation. The theoretical development is illustrated through several nonlinear models of physical and mathematical interest, including equations arising in classical mechanics and nonlinear oscillatory systems. The results demonstrate that commutative factorization constitutes an effective analytical tool for solving nonlinear differential equations.
\end{abstract}

%Uncomment for PACS numbers title message
%\pacs{00.00, 20.00, 42.10}
% Keywords required only for MST, PB, PMB, PM, JOA, JOB?
\vspace{2pc}
%\noindent{\it Keywords}: Planar motion, Rigid body, Fresnel integrals, Jerk
% Uncomment for Submitted to journal title message

%\submitto{\PS}
% Comment out if separate title page not required
\maketitle

\section{Introduction}
The factorization of differential operators has played a fundamental role in the development of mathematical physics for nearly a century. Originally introduced in the context of quantum mechanics by Schrödinger \cite{Schrodinger1940,Schrodinger1941}, and later formalized by Infeld and Hull\cite{InfeldHull1951}, factorization methods have become powerful analytical tools for studying differential equations, constructing exact solutions, and identifying solvable models. Beyond quantum mechanics, these techniques have found numerous applications in nonlinear dynamics, classical mechanics, reaction–diffusion systems, and integrable nonlinear equations.

For nonlinear ordinary differential equations, factorization methods provide an elegant alternative to conventional analytical techniques by reducing a second-order equation into a sequence of first-order differential operators. This decomposition not only simplifies the mathematical treatment of nonlinear models but also establishes direct connections with Riccati and Bernoulli differential equations.
The exact solutions of nonlinear differential equations (ODEs) describe the behavior of a great variety of physical, chemical, biological,
and engineering systems. Widespread systems in these vast areas can be described by Liénard equations, consequently, factorization techniques has become an active research topic in the analysis of nonlinear differential equations.

An efficient factorization procedure for a broad class of second-order nonlinear ordinary differential equations was developed by Rosu and Cornejo-Pérez \cite{RosuCornejo2005a,CornejoRosu2005}. Their method expresses the nonlinear differential operator as the product of two first-order operators, where the factorization functions satisfy a set of nonlinear algebraic and differential relations determined by the coefficients of the original equation. This approach has proven particularly useful for obtaining particular solutions of nonlinear equations arising in classical mechanics, nonlinear oscillations, reaction–diffusion models, and other nonlinear dynamical systems.
Despite its success, the Rosu–Cornejo-Pérez factorization is intrinsically non-commutative. In general, interchanging the order of the first-order operators produces a different differential equation, reflecting the non-commuting nature of the factorization \cite{Gonzalez2024}. Although this property does not affect the validity of the method, it raises a natural mathematical question: under what conditions can the factorizing operators commute while preserving the original nonlinear differential equation? In this paper, the commutative factorization framework for a broad class of second-order nonlinear ordinary differential equations is proposed. We show a systematic route for constructing both particular and general solutions of the original second-order nonlinear equation through a Riccati--Bernoulli equation. Furthermore, a  Bäcklund transformation \cite{LeviBenguria1980,RogersShadwick1982} is derived for the Riccati--Bernoulli equation which allows a known solution to be transformed into a new one. Consequently, the procedure can be iterated to generate an infinite sequence of solutions.\\
The paper is organized as follows. In Section (\ref{sec1}) we briefly review the Rosu–Cornejo-Pérez factorization and we establish the commutative factorization. In Section (\ref{sec2}) we generalize the commutative factorization formalism and derive the route to obtain particular and general solutions of the nonlinear differential equation by solving a Riccati--Bernoulli equation. In Section (\ref{sec3}) a Bäcklund transformation of the Riccati--Bernoulli equation is given. In section (\ref{sec4}) we apply the commutative factorization method to the Sharma-Tasso-Olver-Burgers equation, to the Kundu-Eckhaus equation and the generalized Emden non linear oscillator, respectively. Finally, in the last section we summarize the main results of the commutative factorization method presented in this paper.
\section{Description of the Commutative Factorization}\label{sec1}
Let us consider the following nonlinear second order nonlinear differential equation 
\begin{equation}
\ddot{x} +G(x)\dot{x} + F(x) = 0,\label{eq1}
\end{equation}
where the dot denotes $D=\frac{d}{dt}$ and $G(x)$ and $F(x)$ are arbitrary functions of $x$. Eq. (\ref{eq1}) can be factorized in the form
\begin{equation}\label{eq2}
[ D-\phi_2(x) ] [ D-\phi_1(x) ]x=0 ,
\end{equation}
under the conditions
\begin{eqnarray}
\phi_1+\phi_2+x\frac{d(\phi_1)}{dx}=-G(x), \label{eq3}\\
\phi_1 \phi_2 x =F(x).\label{eq3bis}
\end{eqnarray}
If we interchange the factorization brackets of equation (\ref{eq2}) we get in principle a new second order nonlinear differential equation given by
\begin{equation}
\ddot{x} +g(x)\dot{x} + f(x) = [ D-\phi_1(x) ] [ D-\phi_2(x) ]x=0,\label{eq1a}
\end{equation}
where
\begin{eqnarray}
\phi_1+\phi_2+x\frac{d(\phi_2)}{dx}=-g(x), \label{eq3bi}\\
\phi_1 \phi_2 x =f(x).\label{eq3biss}
\end{eqnarray}
We see that interchanging the factorization brackets implies $f(x)=F(x)$ and $g(x) \neq G(x)$ in general. In order to have $g(x) = G(x)$ when we interchange the factorization functions we need to impose the following condition over the functions $\phi_2 -\phi_1 = -2c_1$ where $c_1$ is an arbitrary real constant. In other words, we have a commutative factorization if and only if the factorization functions differ only by an additive constant.\\
Once a commutative factorization has been found for a given non linear differential equation then we can find the general solution by assuming that $[D-\phi_1(x)]x=\Phi(x,t)$ \cite{Gonzalez2024field},
which yields the following quasi-linear partial differential equation
\begin{equation}
\frac{\partial\Phi}{\partial t} + \frac{\partial\Phi}{\partial x} \left( \Phi + x\phi_1  \right)= \phi_2 \Phi, \label{eq6}
\end{equation}
which can be solved by proposing the ansatz $\Phi(x,t)=x\zeta(t)$ \cite{Gonzalez2025}, which leads to the following Bernoulli equation for the function $\zeta$,
\begin{equation}
\frac{d\zeta(t)}{dt} + \zeta^2(t) = (\phi_2-\phi_1) \zeta(t). \label{eq7}
\end{equation}
If the commutative factorization condition is considered, i.e. $\phi_2 -\phi_1 = -2c_1$ where $c_1$ is a constant, then we get
\begin{equation}
\frac{d\zeta(t)}{dt} + \zeta^2(t) + 2c_1 \zeta(t)=0, \label{eq8}
\end{equation}
One can immediately obtain two particular solutions for (\ref{eq8}) which are $\zeta_{p1}=0$ and $\zeta_{p2}=-2c_1$, which gives us in return two particular solutions of (\ref{eq2}) given by $ [ D-\phi_1(x) ]x=0$ and $ [ D-\phi_2(x) ]x=0$, respectively. This results tells us that if we know a particular solution $\zeta_p$ then we can obtain a  particular solution of the Liénard equation by solving the following first order differential equation $ [ D-\phi_1(x) ]x=x\zeta_p(t)$. On the other hand, if we know a complete solution for $\zeta(t)$, then we can find a complete solution for the non linear differential equation by solving the following first order ODE in the form
\begin{equation}
\dot{x}-\phi_1(x)x=\zeta(t)x. \label{eq10}
\end{equation}
It is worth mentioning that the most general form of a commutative factorization of a given non linear differential equation is given by
\begin{equation}
[D - \varphi_1(x) +b_1 ] [ D- \varphi_1(x) -a_1] x=0 , \label{eq11}
\end{equation}
where $\phi_1(x)= \varphi_1(x) +a_1$ and $\phi_2(x)= \varphi_1(x) -b_1$ such that  $\phi_2 -\phi_1  =-(a_1+b_1)= -2c_1\in \mathbb{R}$, therefore $2c_1=(a_1+b_1)$. By directly calculating (\ref{eq11}) we have :
\begin{equation}\label{eq11a}
\ddot{x} -\dot{x}\left(2\varphi_1(x)+x\frac{d\varphi_1(x)}{dx}+a_1-b_1\right)+x\left(\varphi_1(x)^2+(a_1-b_1)\varphi_1(x)-a_1b_1\right)=0.
\end{equation}
In the next section we will generalize the commutative factorization technique to include higher order non linear terms and show how to obtain a particular and a complete solution of the second order non linear differential equation.

\section{Extension of the Commutative Factorization}\label{sec2}
We will consider now the extension of the commutative factorization of equation (\ref{eq2}) into the following non linear second order differential equation given by
\begin{equation}\label{eq01}
[ D-\varphi_1(x)+b_1 ] [ D-\varphi_1(x)-a_1 ]x=C_0x+C_1\Phi+\frac{C_2}{x}\Phi^2 ,
\end{equation}
where $a_1, b_1 \in \mathbb{R}$, $C_i  \in \mathbb{R}$ for $i=0,1,2$ and $[D-\varphi_1(x)-a_1]x=\Phi(x,t)$. If we assume the following ansatz $\Phi(x,t)=x\zeta(t)^{1-m}$ where $m\neq 1$, we will arrive to the following first order non linear differential equation for $\zeta(t)$ given by
\begin{equation}\label{eq03}
\frac{d\zeta(t)}{dt} =a\zeta(t)^{2-m}+b\zeta(t)+c\zeta^m(t)
\end{equation}
where $a=(C_2-1)/(1-m)$, $b=(C_1-2c_1)/(1-m)$ and $c=C_0/(1-m)$.  Equation (\ref{eq03}) is known as a Riccati-Bernoulli equation \cite{Yang2015,Mirzazadeh2017,Hassan2019,Abdelrahman2018,Alharbi2020b}. When $ac\neq 0$ and $m=0$ we have a Riccati equation. When $a\neq 0$ and $c=0$ we have a Bernoulli equation. Equation (\ref{eq03}) has solutions as follows:
\begin{description}
  \item[Case ] 1. When $a\neq 0$ and $4(C_2-1)C_0-(C_1-2c_1)^2>0$ a complete solution of equation (\ref{eq03}) is
   \begin{align}
  % \zeta(t)^{1-m}&=-\frac{C_1-2c}{2(C_2-1)}+\frac{\sqrt{4(C_2-1)C_0-(C_1-2c)^2}}{2(C_2-1)}\tan[\frac{\sqrt{4(C_2-1)C_0-(C_1-2c)^2}}{2}(t+\delta)] \label{eq04}\\
\zeta(t)^{1-m}&=-\frac{C_1-2c_1}{2(C_2-1)}+\frac{\sqrt{4(C_2-1)C_0-(C_1-2c_1)^2}}{2(C_2-1)}\tan[\frac{\sqrt{4(C_2-1)C_0-(C_1-2c_1)^2}}{2}(t+\delta)] \label{eq04}\\
\zeta(t)^{1-m}&=-\frac{C_1-2c_1}{2(C_2-1)}-\frac{\sqrt{4(C_2-1)C_0-(C_1-2c_1)^2}}{2(C_2-1)}\cot[\frac{\sqrt{4(C_2-1)C_0-(C_1-2c_1)^2}}{2}(t+\delta)] \label{eq04a}
\end{align}
  \item[Case] 2. When $a\neq 0$ and $4(C_2-1)C_0-(C_1-2c_1)^2<0$ the particular and a complete solution of equation (\ref{eq03}) is
  \begin{align}
\zeta_{p\pm}(t)^{1-m}&=\frac{-(C_1-2c_1)\pm\sqrt{(C_1-2c_1)^2-4(C_2-1)C_0}}{2(C_2-1)}  \label{eq05a}\\
\zeta(t)^{1-m}&=-\frac{C_1-2c_1}{2(C_2-1)}-\frac{\sqrt{(C_1-2c_1)^2-4(C_2-1)C_0}}{2(C_2-1)}\tanh[\frac{\sqrt{(C_1-2c_1)^2-4(C_2-1)C_0}}{2}(t+\delta)] \label{eq05}\\
\zeta(t)^{1-m}&=-\frac{C_1-2c_1}{2(C_2-1)}-\frac{\sqrt{(C_1-2c_1)^2-4(C_2-1)C_0}}{2(C_2-1)}\coth[\frac{\sqrt{(C_1-2c_1)^2-4(C_2-1)C_0}}{2}(t+\delta)] \label{eq05c}
\end{align}
  \item[Case] 3. When $a\neq 0$ and $4(C_2-1)C_0-(C_1-2c_1)^2=0$ the particular and a complete solution of equation (\ref{eq03}) is
   \begin{align}
   \zeta_{p}(t)^{1-m}&=-\frac{C_1-2c_1}{2(C_2-1)} \label{eq06a}\\
\zeta(t)^{1-m}&=-\frac{C_1-2c_1}{2(C_2-1)}-\frac{1}{(C_2-1)(t+\delta)}\label{eq06}
\end{align}
  \item[Case] 4. When $a\neq 0$, $b\neq 0$ and $C_0=0$ a complete solution of equation (\ref{eq03}) is
   \begin{align}
    \zeta_{p}(t)^{1-m}&=0\label{eq07a}\\
  \zeta_{p}(t)^{1-m}&=-\frac{C_1-2c_1}{C_2-1}\label{eq07b}\\
\zeta(t)^{1-m}&=\delta e^{(C_1-2c_1)t}-\frac{C_2-1}{(C_1-2c_1)}\label{eq07}
\end{align}
\item[Case] 5. When $a\neq 0$, $C_0=0$ and $C_1=2c_1$ the particular and a complete solution of equation (\ref{eq03}) is
  \begin{align}
  \zeta_{p}(t)^{1-m}&=0\label{eq08a}\\
\zeta(t)^{1-m}&=-\frac{1}{(C_2-1)(t+\delta)}\label{eq08}
\end{align}
\end{description}
where $\delta$ is an arbitrary constant of integration. Therefore, by solving the following first order non linear differential equation
\begin{equation}\label{eq09}
 [ D-\varphi_1(x)-a_1 ]x=x\zeta(t)^{1-m}
\end{equation}
we arrive to the general solution of the second order non linear differential equation given by
\begin{equation}\label{eq010}
[ D-\varphi_1(x)+b_1 ] [ D-\varphi_1(x)-a_1 ]x=C_0x+C_1(\dot{x}-x\varphi_1(x)-a_1)+\frac{C_2}{x}(\dot{x}-x\varphi_1(x)-a_1)^2.
\end{equation}
In the subsequent section we will give a Bäcklund transformation of the Riccati-Bernoulli equation that will allow us to obtain a sequence of solutions for equation (\ref{eq010}).
\section{Bäcklund transformation of the Riccati-Bernoulli equation}\label{sec3}
Let $\zeta_{n-1}(t)$ and $\zeta_n(t)$, with
\[
\zeta_n(t)=\zeta_n\left(\zeta_{n-1}(t)\right),
\]
be solutions of Eq.~(\ref{eq03}), then
\begin{equation}
\frac{d\zeta_n(t)}{dt}
=
\frac{d\zeta_n(t)}{d\zeta_{n-1}(t)}
\frac{d\zeta_{n-1}(t)}{dt}
=
\frac{d\zeta_n(t)}{d\zeta_{n-1}(t)}
\left(
a \zeta_{n-1}^{2-m}
+b \zeta_{n-1}
+c \zeta_{n-1}^{m}
\right).
\end{equation}
Thus,
\begin{equation}
\frac{d\zeta_n(t)}
{a \zeta_n^{2-m}+b \zeta_n+c \zeta_n^m}
=
\frac{d\zeta_{n-1}(t)}
{a \zeta_{n-1}^{2-m}+b \zeta_{n-1}+c \zeta_{n-1}^m}.
\end{equation}

Integrating the above equation once with respect to $t$ and
simplifying, we obtain

\begin{equation}\label{eq13}
\zeta_n(t)^{1-m}
=
\frac{
-cA_1+aA_2
\left(\zeta_{n-1}(\xi)\right)^{1-m}
}{
bA_1+aA_2+aA_1
\left(\zeta_{n-1}(\xi)\right)^{1-m}
},
\end{equation}
where $A_1$ and $A_2$ are arbitrary constants.

Equation~(\ref{eq13}) represents a Bäcklund transformation for Eq.~(\ref{eq03}).
Therefore, once a solution of Eq.~(\ref{eq03}) is known, Eq.~(\ref{eq13}) can be used
iteratively to construct an infinite sequence of solutions of
Eq.~(\ref{eq03}). Consequently, an infinite sequence of solutions of Eq.~(\ref{eq010})
can also be obtained  by solving the following first order non linear differential equation
\begin{equation}\label{eq14}
 [ D-\varphi_1(x)-a_1 ]x=x\zeta_{n}(t)^{1-m}.
\end{equation}
In the subsequent section we will give some examples to illustrate the utility of the extension of the commutative factorization method.
\section{Examples}\label{sec4}

\subsection{Example 1. Sharma-Tasso-Olver-Burgers Equation}
\begin{figure}[ht]
\includegraphics[width=0.8\textwidth]{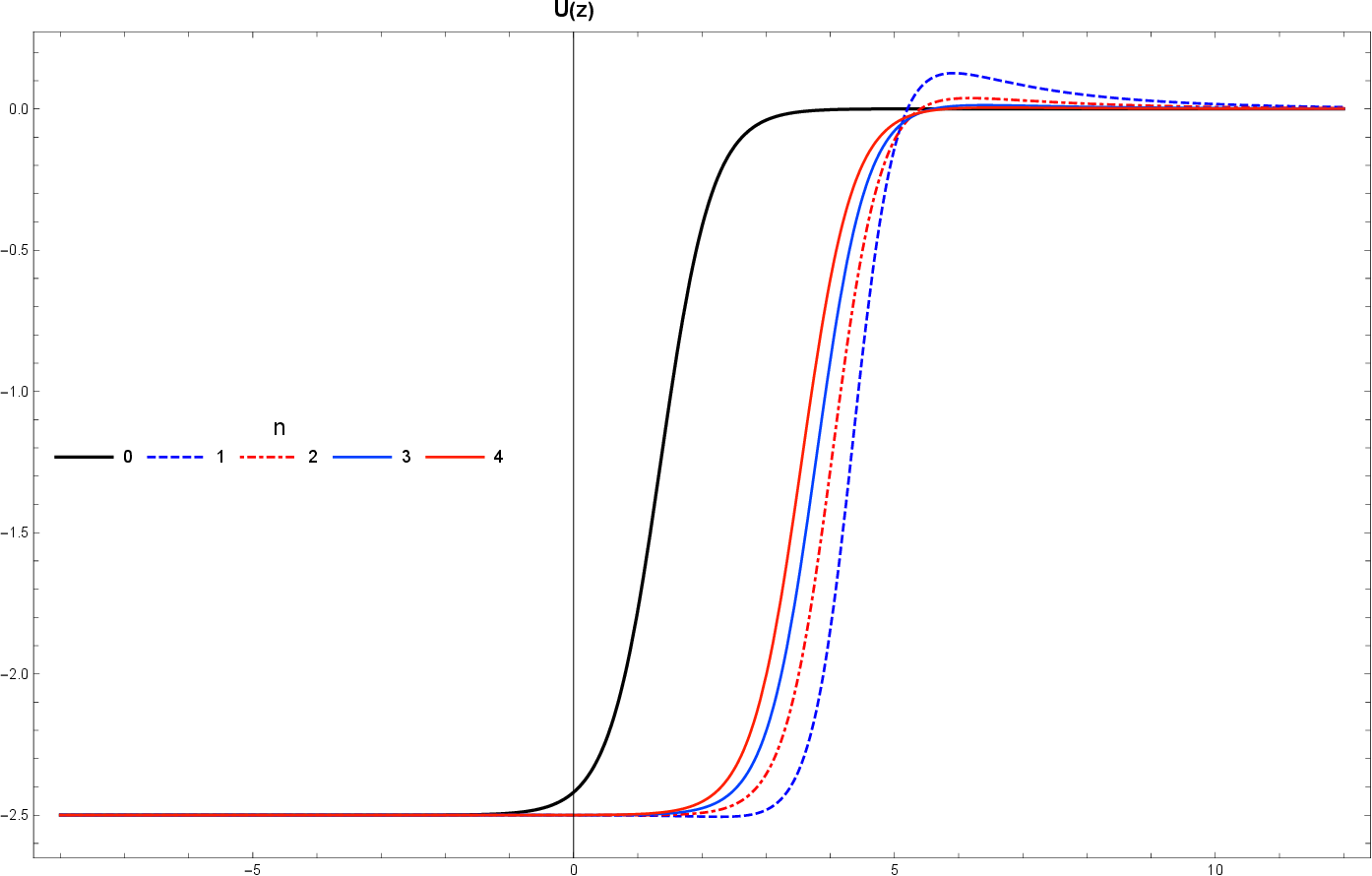}
\caption{Illustration of the traveling wave solutions of the STOB equation obtained by using the  Bäcklund transformation. We use for these curves the following values for the constants : $A_1=-0.1$, $A_2=0.5$, $\delta=-5$, $\alpha=1$, $\beta=1$, $v=-1.25$ and $\lambda=1$.}
\label{stobfig}
\end{figure}

We will first study the Sharma-Tasso-Olver-Burgers (STOB) equation which is given by
\begin{equation}
u_t+\alpha(3u_x^{2}+3u^2u_x+3uu_{xx}+u_{xxx})+\beta(2uu_x+u_{xx})= 0.\label{eq22}
\end{equation}
Clearly, it is the Burgers equation when $\alpha= 0$, and it reduces to the Sharma–Tasso–Olver
equation when $\beta = 0$. The STOB equation is a nonlinear evolution equation that combines the dispersive and nonlinear structure of the Sharma–Tasso–Olver (STO) equation with the dissipative characteristics of the Burgers equation. This combination makes the STOB equation an interesting model for investigating the interplay between nonlinear dispersion, diffusion, and wave propagation in nonlinear dynamical systems \cite{Miao2021,KaiYin2022,ZhouZhuang2022}.
This equation incorporates both dissipative and higher-order nonlinear effects, providing a richer dynamical structure than either of its constituent equations. To study the traveling wave solutions of equation (\ref{eq22}) we make the following change of variables $u=U(z)$ where $z=x-vt$ and substitute into equation (\ref{eq22}). Now, integrating once and letting the integration constant to be zero, we arrive to the following second order non linear differential equation
\begin{equation}
U^{\prime\prime}+(3U+ \frac{\beta}{\alpha})U^{\prime}+U(U^2+ \frac{\beta}{\alpha}U- \frac{v}{\alpha})= 0, \label{eq23}
\end{equation}
where $U^{\prime}=\frac{dU}{dz}$. Suppose that we want to write down equation (\ref{eq23}) in the factorization form given by equation (\ref{eq010}); to do this we need first to expand equation (\ref{eq010}) and compare it with equation (\ref{eq23}) in order to set an algebraic system of equations by comparing the coefficients of both equations, which for this case is given by
\begin{align}
  2\varphi_1+U\frac{d\varphi_1}{dU}-(a_1-b_1) &=-3U -\frac{\beta}{\alpha} \\
 \varphi_1^2+(a_1-b_1) \varphi_1- a_1b_1-C_0 &=- \frac{v}{\alpha}
\end{align}
where we have taken $C_1=C_2=0$. By setting $\varphi_1=-U$ and $a_1=0$ into the above equations we have the following solution for the coefficients $b_1=2c_1=\beta/\alpha$ and $C_0=v/\alpha$. Another possible solution is to set $\varphi_1=-U-\beta/2\alpha$ and $a_1=b_1=c_1=0$ and $C_0=(\beta/2\alpha)^2+v/\alpha$.
Then, one can factorize equation (\ref{eq23}) in the following two different ways
\begin{align}
  \left[ D+U+\frac{\beta}{\alpha}\right] \left[ D+U \right]U &= \frac{v}{\alpha}U  \label{eq24a}\\
  \left[ D+U+\frac{\beta}{2\alpha}\right]^2 U &= \left[\left(\frac{\beta}{2\alpha}\right)^2+\frac{v}{\alpha}\right]U.  \label{eq24b}
\end{align}
Both of the above equations give the same result, so we are going to work with equation (\ref{eq24b}) without loss of generality. We are going also to assume that $\alpha>0$, $\beta\geq0$ and $v<0$, such that $C_0=(\beta/2\alpha)^2+v/\alpha$ can take any possible real value. The solution for equation (\ref{eq24b}) is obtained by solving the following Bernoulli equation
\begin{equation}\label{eq24c}
 [ D+U+\frac{\beta}{2\alpha} ]U=U\zeta_{n}(z)^{1-m}.
\end{equation}
Equation (\ref{eq24c}) has solutions as follows:
\begin{enumerate}
  \item When $-4C_0=-4(\frac{\beta^2}{4\alpha^2}+\frac{v}{\alpha})>0$, the solution of equation (\ref{eq24b}) is
  \begin{align}
   U(z)& = \frac{4 \alpha ^2 C_0-\beta ^2}{2 \alpha  \left(\beta -2 \alpha  \sqrt{-C_0} \tan \left(\sqrt{-C_0} z\right)\right)-\lambda \left(\beta ^2-4 \alpha ^2 C_0\right) \sec \left(\sqrt{-C_0} z\right) e^{\frac{\beta
   z}{2 \alpha }}}\label{eq25}\\
   U(z)& =\frac{\left(4 \alpha ^2 C_0-\beta ^2\right) \sin \left(\sqrt{-C_0} z\right)}{-\lambda \left(\beta ^2-4 \alpha ^2 C_0\right) e^{\frac{\beta  z}{2 \alpha }}+4 \alpha ^2 \sqrt{-C_0} \cos \left(\sqrt{-C_0}
   z\right)+2 \alpha  \beta  \sin \left(\sqrt{-C_0} z\right)}\label{eq25m}
  \end{align}
  \item When $-4C_0<0$, the solution of equation (\ref{eq24b}) is
  \begin{align}
  U_{p\pm}(z)&=\frac{\beta \pm2 \alpha  \sqrt{C_0}}{e^{\left(\frac{\beta }{2 \alpha }\pm\sqrt{C_0}\right) (z-2 \alpha  \delta )}-2 \alpha }\label{eq26a}\\
   U(z) &= \frac{\left(4 \alpha ^2 C_0-\beta ^2\right) \cosh \left(\sqrt{C_0} z\right)}{-\lambda \left(\beta ^2-4 \alpha ^2 C_0\right) e^{\frac{\beta  z}{2 \alpha }}+4 \alpha ^2 \sqrt{C_0} \sinh \left(\sqrt{C_0} z\right)+2
   \alpha  \beta  \cosh \left(\sqrt{C_0} z\right)}\label{eq26}\\
   U(z)&=\frac{\left(4 \alpha ^2 C_0-\beta ^2\right) \sinh \left(\sqrt{C_0} z\right)}{-\lambda \left(\beta ^2-4 \alpha ^2 C_0\right) e^{\frac{\beta  z}{2 \alpha }}+4 \alpha ^2 \sqrt{C_0} \cosh \left(\sqrt{C_0} z\right)+2
   \alpha  \beta  \sinh \left(\sqrt{C_0} z\right)}\label{eq26b}
  \end{align}
  \item When $C_0=0$, the solution of equation (\ref{eq24b}) is
 \begin{align}
  U_p(z)&=\frac{\beta }{2 \alpha -e^{\frac{\beta  z}{2 \alpha }-\beta  \delta }}\label{eq27a}\\
   U(z)& = \frac{\beta ^2 (\delta +z)}{-4 \alpha ^2+\beta ^2 \lambda e^{\frac{\beta
    z}{2 \alpha }}-2 \alpha  \beta  (\delta +z)}\label{eq27}
 \end{align}
\end{enumerate}
Now we will apply the Bäcklund transformation to get new solutions. We will be interested in kink traveling solutions of the STOB equation, therefore we will be interested in the case when $-4C_0<0$. Using as our seed solution the function $\zeta_{0}(z)^{1-m}=\sqrt{C_0} \coth \left(\sqrt{C_0} (z+\delta)\right)$ obtained through equation (\ref{eq05c}) and using equation (\ref{eq13}) we find another solution for the Riccati-Bernoulli equation given by
\begin{equation}\label{eq28}
 \zeta_{1}(z)^{1-m}= \frac{-A_1 C_0-A_2 \sqrt{C_0} \coth \left(\sqrt{C_0} (z+\delta)\right)}{-A_1 \sqrt{C_0} \coth \left(\sqrt{C_0} (z+\delta)\right)-A_2}.
\end{equation}
By inserting equation (\ref{eq28}) into equation (\ref{eq24c}) and solving the Bernoulli equation we find a new solution for the STOB equation given by
\begin{equation}\label{eq29}
  U(z) =\frac{\left(4 \alpha ^2 C_0-\beta ^2\right) \left(A_2 \sinh \left(\sqrt{C_0} (\delta +z)\right)+A_1 \sqrt{C_0} \cosh \left(\sqrt{C_0} (\delta +z)\right)\right)}{2 \alpha  \left(A_2 \beta +2 \alpha  A_1
   C_0\right) \sinh \left(\sqrt{C_0} (\delta +z)\right)+2 \alpha  \sqrt{C_0} \left(2 \alpha  A_2+A_1 \beta \right) \cosh \left(\sqrt{C_0} (\delta +z)\right)-\lambda e^{\frac{\beta  (\delta +z)}{2 \alpha }}\left(\beta ^2-4 \alpha ^2 C_0\right)}
\end{equation}
where $A_1$, $A_2$, $\delta$ and $\lambda$ are arbitrary constants. We can apply this procedure an infinite number of times and obtain an infinite sequence of solutions. In figure (\ref{stobfig}) we show the plots of the solution of the STOB equation using the Bäcklund transformation for $n=0,1,2,3,4$.

\subsection{Example 2. Kundu-Eckhaus Equation}
The Kundu–Eckhaus equation (KEE) is an important integrable nonlinear Schrödinger-type equation that appears in the study of nonlinear wave propagation, particularly when the dynamics include an additional nonlinear phase modulation or derivative-type interaction \cite{Kundu1984}. In particular, it provides an extension of the cubic nonlinear Schrödinger equation by incorporating quintic nonlinear and intensity-gradient terms \cite{CalogeroEckhaus1987}. These additional contributions account for higher-order nonlinear phase effects and make the KEE relevant to a variety of physical applications, including nonlinear optics, plasma physics, and Bose–Einstein condensates \cite{Wang2014,Qiu2015,CuiWang2023}. In nonlinear optical systems, the equation can be used to describe pulse propagation in regimes where the standard Kerr nonlinear Schrödinger model requires higher-order nonlinear corrections. The derivative nonlinear term introduces an intensity-dependent phase modulation associated with spatial variations of the wave intensity. A commonly used form of the Kundu–Eckhaus equation is
\begin{equation}\label{eq30}
i \Psi_t+\Psi_{xx}+2 (|\Psi|^2)_x\Psi+|\Psi|^4\Psi=0,
\end{equation}
where $\Psi=\Psi(x,t)$ is a complex-valued function of two real variables. Using the traveling wave transformation $\Psi(x,t)=u(\xi)e^{i(\alpha x+\beta t)}$ in equation (\ref{eq30}) we obtain
\begin{equation}\label{eq31}
  u^{\prime\prime}+\frac{4u^2}{k}u^{\prime}- \frac{\alpha^2+\beta}{k^2}u+ \frac{u^5}{k^2}= 0,
\end{equation}
where $\xi= k(x-2\alpha t)$, and $k$, $\alpha$ and $\beta$ are real constants.
 Suppose that we want to write down equation (\ref{eq31}) in the factorization form given by equation (\ref{eq010}); to do this we need first to expand equation (\ref{eq010}) and compare it with equation (\ref{eq31}) in order to set an algebraic system of equations by comparing the coefficients of both equations, which for this case is given by
\begin{align}
  2\varphi_1+u\frac{d\varphi_1}{du}-(a_1-b_1) &= -\frac{4u^2}{k} \\
 \varphi_1^2+(a_1-b_1) \varphi_1- a_1b_1-C_0 &=- \frac{\alpha^2+\beta}{k^2}+ \frac{u^4}{k^2}
\end{align}
where we have taken $C_1=C_2=0$. If $a_1=b_1=c_1=0$ then $\varphi_1(u)=u^2/k$ and $C_0=(\alpha^2+\beta)/k^2$. Then, one can factorize equation (\ref{eq31}) in the following way
\begin{equation} \label{eq32}
  \left[ D+\frac{u^2}{k}\right]^2 u =\left(\frac{\alpha^2+\beta}{k^2}\right)u.
\end{equation}
The solution for equation (\ref{eq32}) is obtained by solving the following Bernoulli equation
\begin{equation}\label{eq33}
\left[ D+\frac{u^2}{k}\right]u=u\zeta_{n}(\xi)^{1-m}.
\end{equation}
Equation (\ref{eq33}) has solutions as follows:
\begin{enumerate}
  \item When $-4C_0=-4((\alpha^2+\beta)/k^2)>0$, the solution of equation (\ref{eq32}) is
  \begin{align}
   u(\xi)& = \pm\frac{\sqrt{2} \sqrt[4]{-C_0} \sqrt{k}}{\sqrt{\sec ^2\left(\sqrt{-C_0} (\xi+\delta)\right) \left(\sin \left(2 \sqrt{-C_0} (\xi+\delta)\right)+2 \sqrt{-C_0} (\xi+\delta+\lambda k)\right)}}\label{eq34}\\
   u(\xi)& =\pm\frac{\sqrt{2} \sqrt[4]{-C_0} \sqrt{k}}{\sqrt{\csc ^2\left(\sqrt{-C_0} (\xi+\delta)\right) \left(-\sin \left(2 \sqrt{-C_0} (\xi+\delta)\right)+2 \sqrt{-C_0} (\xi+\delta+\lambda k)\right)}}\label{eq35}
  \end{align}
  \item When $-4C_0<0$, the solution of equation (\ref{eq33}) is
  \begin{align}
  u_{p\pm}(\xi)&=\frac{\pm\sqrt{k} \sqrt[4]{C_0}}{\sqrt{\pm1\mp e^{\mp2\sqrt{C_0}(\xi\pm\lambda k)}}}\label{eq36a}\\
   u(\xi) &= \pm\frac{\sqrt{2} \sqrt[4]{C_0} \sqrt{k}}{\sqrt{\text{sech}^2\left(\sqrt{C_0} (\xi+\delta)\right) \left(\sinh \left(2 \sqrt{C_0} (\xi+\delta)\right)+2 \sqrt{C_0} (\xi+\delta+\lambda k)\right)}}\label{eq36}\\
   u(\xi)&=\pm\frac{\sqrt{2} \sqrt[4]{C_0} \sqrt{k}}{\sqrt{\text{csch}^2\left(\sqrt{C_0} (\xi+\delta)\right) \left(\sinh \left(2 \sqrt{C_0} (\xi+\delta)\right)-2 \sqrt{C_0} (\xi+\delta-\lambda k)\right)}}\label{eq36b}
  \end{align}
  \item When $C_0=0$, the solution of equation (\ref{eq33}) is
 \begin{equation}
  u_p(\xi)=\sqrt{\frac{k}{2(\xi+\lambda)}}\label{eq37}\\
 \end{equation}
\end{enumerate}
Now we will apply the Bäcklund transformation to get new solutions. Using as our seed solution the function $\zeta_{0}(\xi)^{1-m}=\sqrt{C_0} \coth \left(\sqrt{C_0} (\xi+\delta)\right)$ obtained through equation (\ref{eq05c}) and using equation (\ref{eq13}) we find another solution for the Riccati-Bernoulli equation given by
\begin{equation}\label{eq38}
 \zeta_{1}(\xi)^{1-m}= \frac{-A_1 C_0-A_2 \sqrt{C_0} \coth \left(\sqrt{C_0} (\xi+\delta)\right)}{-A_1 \sqrt{C_0} \coth \left(\sqrt{C_0} (\xi+\delta)\right)-A_2}.
\end{equation}
By inserting equation (\ref{eq38}) into equation (\ref{eq33}) and solving the Bernoulli equation we find a new solution for the KEE equation given by
\begin{equation}\label{39}
  u(\xi) =\frac{\sqrt{2} \sqrt{-\sqrt{C_0} k \left(A_1 \sqrt{C_0} \cosh \left(\sqrt{C_0} (\xi+\delta)\right)+A_2 \sinh \left(\sqrt{C_0} (\xi+\delta)\right)\right)^2}}{\sqrt{-2 \sqrt{C_0} \left(A_1^2 C_0
   (\xi+\delta)-A_2^2 (\xi+\delta)+\lambda k\right)-\left(A_1^2 C_0+A_2^2\right) \sinh \left(2 \sqrt{C_0} (\xi+\delta)\right)-2 A_1 A_2 \sqrt{C_0} \cosh \left(2 \sqrt{C_0} (\xi+\delta)\right)}}
\end{equation}
where $A_1$, $A_2$, $\delta$ and $\lambda$ are arbitrary constants. We can get an infinite sequence of solutions by applying this process over and over again.

\subsection{Example 3. Isochronous generalized Emden oscillator}
\begin{figure}[ht]
\includegraphics[width=0.8\textwidth]{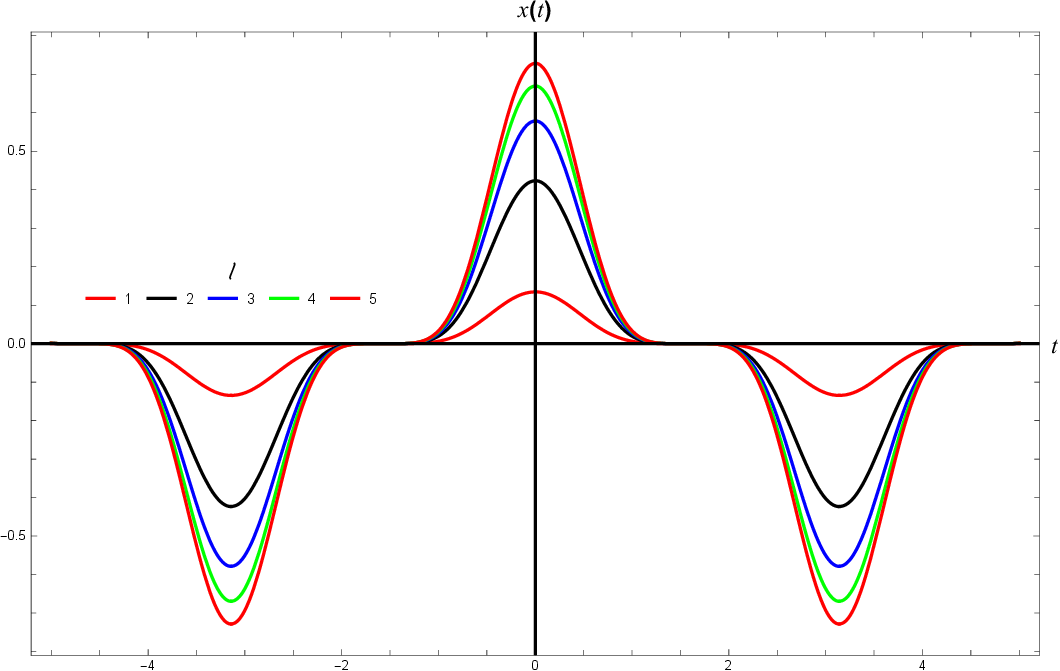}
\caption{Illustration of the solutions of the Emden equation for the following values for the constants : $\omega=1$, $k=1$, $\delta=0$, $\lambda=400$, $m=2$ and $l=1,2,3,4,5$.}
\label{emfig}
\end{figure}

We will study next a generalization of the Emden class of non linear oscillators which is obtained by choosing
\begin{equation}
\phi_1(x)=\frac{C_1-2c_1}{2(C_2-1)}-kx^{q} \quad \mbox{and}  \quad \phi_2(x)=\frac{C_1-2c_1}{2(C_2-1)}-kx^{q}-2c_1, \label{eq14}
\end{equation}
where $q \geq 0$ is a positive integer. Note that $(\phi_2-\phi_1)=-2c_1$. Substituting equation (\ref{eq14}) into equation (\ref{eq010}) we obtain the following non linear differential equation
\begin{equation}
\ddot{x}-k\left[2(C_2-1)-q\right]x^{q}\dot{x}-\frac{4(C_2-1)C_0-(C_1-2c)^2}{4(C_2-1)}x+k^2(1-C_2)x^{2q+1} =\frac{C_2}{x}\dot{x}^2 . \label{eq15}
\end{equation}
The importance of the study of equation (\ref{eq15}) arises from the fact that this equation contains many physically interesting equations
such as the modified Emden equation, unforced Duffing oscillator, Helmholtz oscillator, etc and is intimately related to the two dimensional
Lotka-Volterra system \cite{Pradeep2010}.
If we restrict ourselves to the case when $4(C_2-1)C_0-(C_1-2c_1)^2>0$, then equation (\ref{eq15}) represents a class of solvable nonlinear oscillators with periodic solutions. In particular, by choosing in Eq. (\ref{eq15}) $k=2m+1$, $C_0=-(2m+1)$, $C_1=2c_1$, $C_2=2m/(2m+1)$ and $q=1$ where $m$ is a non negative integer we obtain the system studied in Ref. (\cite{Gonzalez2025parametric}).  \\
The general solution to equation (\ref{eq15}) is obtained by substituting $\phi_1(x)=(C_1-2c_1)/2(C_2-1)-kx^{q}$ into equation (\ref{eq09}) and using $\zeta(t)$ from equation (\ref{eq04}) to obtain the following Bernoulli differential equation
\begin{equation}
\dot{x}-\frac{\sqrt{4(C_2-1)C_0-(C_1-2c_1)^2}}{2(C_2-1)}\tan[\frac{\sqrt{4(C_2-1)C_0-(C_1-2c_1)^2}}{2}(t+\delta)]x =-kx^{q+1} . \label{eq16}
\end{equation}
The solution to equation (\ref{eq16}) is given by
\begin{equation}\label{eq17}
  x(t)=\frac{\cos^{1/(1-C_2)}[\omega(t+\delta)]}{\left(\lambda+k q\int^{t}\cos^{q/(1-C_2)}[\omega(t^{\prime}+\delta)]dt^{\prime}\right)^{1/q}}.
\end{equation}
where $\omega=\sqrt{4(C_2-1)C_0-(C_1-2c_1)^2}/2$ and $\lambda$ is an integration constant. In order to obtain isochronous orbits we need to define $q=2l+1$ and $1/(1-C_2)=2m+1$ where $l$ and $m$ are non negative integers. The solution therefore is given by:
\begin{equation}\label{eq18}
  x(t)=\frac{\cos^{(2m+1)}[\omega(t+\delta)]}{\left[\lambda-\frac{k (2l+1)}{2N \omega}\left(\cos ^{2 N}(\omega(t+\delta)) \, _2F_1\left(\frac{1}{2},N;N+1;\cos ^2(\omega(t+\delta))\right)\right)\right]^{2l+1}}
\end{equation}
where $N=2 m l+m+l+1$ and $_2F_1(a,b;c;z)$ is the Hypergeometric function. It is necessary to choose the constant of integration $\lambda$ in such a way as to avoid a singularity of the solution given in equation (\ref{eq17}) in the denominator. In figure (\ref{emfig}) we show the solution of the generalized Emden oscillator

\section{Conclusions}

In this work, we have developed a commutative factorization framework for a broad class of nonlinear second-order ordinary differential equations. The main objective was to extend the commutative factorization procedure in order to use the Riccati-Bernoulli equation and the Bäcklund transformation. This provides a systematic route for constructing both particular and general solutions of the original second-order nonlinear equation. Furthermore, the  Bäcklund transformation derived for the Riccati--Bernoulli equation allows a known solution to be transformed into a new one. Consequently, the procedure can be iterated to generate an infinite sequence of solutions.

The effectiveness of the proposed factorization is illustrated through three representative examples. First, the traveling-wave reduction of the Sharma--Tasso--Olver--Burgers (STOB) equation is transformed into a second-order nonlinear ordinary differential equation that admits the proposed commutative factorization. The resulting solutions include hyperbolic and trigonometric traveling-wave structures, and the  Bäcklund transformation provides an iterative mechanism for constructing additional solutions. This result is consistent with the rich structure of exact solutions previously reported for the STOB equation.

Second, the method is applied to the Kundu--Eckhaus equation. After a traveling-wave reduction, the resulting nonlinear ordinary differential equation is factorized in a commutative form, leading to explicit solutions expressed in terms of trigonometric and hyperbolic functions. The construction also generates an infinite sequence of solutions through repeated application of the  Bäcklund transformation. This result is particularly relevant because the Kundu--Eckhaus equation is an integrable nonlinear Schr\"odinger-type model with important connections to gauge transformations and nonlinear wave propagation.

Finally, we considered a generalized Emden-type oscillator. The resulting class of nonlinear equations contains several physically and mathematically relevant models, including generalized Emden, Duffing-type, and Helmholtz-type oscillators. For appropriate parameter choices, the factorization leads to periodic and isochronous solutions. This example demonstrates that the commutative factorization is not restricted to traveling-wave equations but can also be applied to nonlinear oscillatory systems.

Overall, the results indicate that commutative factorization provides a useful complementary approach to existing techniques for solving nonlinear differential equations. Its main advantage is that the commutativity condition produces a family of equivalent factorizations and establishes a direct connection between nonlinear second-order equations, Riccati--Bernoulli equations, and  Bäcklund transformation. In this sense, the method provides not only a procedure for obtaining exact solutions but also a framework for investigating the algebraic structure underlying nonlinear dynamical systems.

\end{document}